\documentclass[twocolumn,amsmath,amssymb]{revtex4}

\usepackage{graphicx}
\usepackage{graphics}
\usepackage{subfigure}
\usepackage{dcolumn}
\usepackage{bm}
\usepackage{txfonts}

\begin{document}

\title{Social Imitation Dynamics of Vaccination Driven by Vaccine Effectiveness and Beliefs}
\author{Feng Fu$^{1,2}$}
\author{Ran Zhuo$^{1}$}
\author{Xingru Chen$^{3}$}
\email{xingrucz@gmail.com}

\affiliation{ $^1$Department of Mathematics, Dartmouth College, Hanover, NH 03755, USA\\
$^2$Department of Biomedical Data Science, Geisel School of Medicine at Dartmouth, Lebanon, NH 03756, USA\\
$^3$Department of Mathematics, Beijing University of Posts and Telecommunications, Beijing 100876, China}

\date{\today}

\begin{abstract}
Declines in vaccination coverage for vaccine-preventable diseases, such as measles and chickenpox, have enabled their surprising comebacks and pose significant public health challenges in the wake of growing vaccine hesitancy. Vaccine opt-outs and refusals are often fueled by beliefs concerning perceptions of vaccine effectiveness and exaggerated risks. Here, we quantify the impact of competing beliefs -- vaccine-averse versus vaccine-neutral -- on social imitation dynamics of vaccination, alongside the epidemiological dynamics of disease transmission. These beliefs may be pre-existing and fixed, or coevolving attitudes. This interplay among beliefs, behaviors, and disease dynamics demonstrates that individuals are not perfectly rational; rather, they base their vaccine uptake decisions on beliefs, personal experiences, and social influences. We find that the presence of a small proportion of fixed vaccine-averse beliefs can significantly exacerbate the vaccination dilemma, making the tipping point in the hysteresis loop more sensitive to changes in individuals' perceived costs of vaccination and vaccine effectiveness. However, in scenarios where competing beliefs spread concurrently with vaccination behavior, their double-edged impact can lead to self-correction and alignment between vaccine beliefs and behaviors. The results show that coevolution of vaccine beliefs and behaviors makes populations more sensitive to abrupt changes in perceptions of vaccine cost and effectiveness compared to scenarios without beliefs. Our work provides valuable insights into harnessing the social contagion of even vaccine-neutral attitudes to overcome vaccine hesitancy.
\end{abstract}

\keywords{Vaccine hesitancy, Vaccination dilemma, Hysteresis, Behavioral Epidemiology} 
\maketitle

\section*{Introduction}

Mass vaccination is one of the most cost-effective pharmaceutical interventions for the prevention and control of infectious diseases~\cite{anderson1991infectious,pastor2002immunization}. Long before the advent of modern vaccine technology, people in China, around 1000 AD, used pus from cows infected with cowpox to inoculate humans against smallpox~\cite{needham1980china}. Fast forward to today, the worldwide deployment of modern vaccines has drastically reduced morbidity and mortality, particularly for childhood diseases. Despite the fact that vaccines have saved millions of lives over decades---most notably with the record-breaking innovation of mRNA vaccines during the recent COVID-19 pandemic~\cite{kabir2020evolutionary,siegler2021trajectory,oanua2024online}---achieving widespread population immunity through voluntary vaccination remains a persistent public health challenge~\cite{wadman2017vaccine}.

In recent years, for example, the world has seen a resurgence of vaccine-preventable diseases, such as measles~\cite{jansen2003measles}, polio~\cite{tediosi2024leveraging} and pertussis (whooping cough)~\cite{asch1994omission}. Notably, measles has recently reached the very verge of an endemic disease in France~\cite{jansen2003measles}. Polio, once on the brink of global eradication, remains endemic in Afghanistan and Pakistan~\cite{tediosi2024leveraging}. Whooping cough is on the rapid rise in North America and Europe~\cite{smout2024whooping} 

Such resurgent outbreaks of measles and other diseases suggest substandard vaccination compliance~\cite{majumder2015substandard}, despite tremendous efforts to address vaccine hesitancy~\cite{antona2013measles,gowda2013rise}, especially given how cost-effective these vaccines actually are. For instance, prior study estimates that, as compared to the cost of one measles vaccine $\$20$, the cost to treat each measles infection is $\$10,376$, while the total cost to contain each outbreak is $\$124,517$~\cite{sugerman2010measles}. Even though resurgent measles outbreaks impose huge risks for those unvaccinated and even in some European regions it has become an endemic disease~\cite{jansen2003measles,antona2013measles}, the coverage of measles vaccination still remains insufficient~\cite{majumder2015substandard}, for more than a decade following the infamous MMR vaccination and autism controversy~\cite{burgess2006mmr}.

Notably in the aftermath of a sharp decline in vaccination coverage triggered by concerns regarding vaccine safety and efficacy, the recovery of vaccination rate from nadir to levels needed to attain herd immunity has been remarkably slow. For example, it takes almost two decades for the recovery in the uptake of whole-cell pertussis vaccine from rock bottom 30\% in 1978 to 91\% in 1992 in England and Wales~\cite{miller1992epidemiology,baker2003pertussis,rohani2000pertussis}. This suggests that the recovery of vaccination rate depends not just on the extent of mitigating perceived cost of vaccination and improving vaccine efficacy, but also on the past vaccination trajectory, hence possibly hindering a rapid increase. To shed light on this puzzling phenomenon, recent theoretical study finds that social imitation dynamics of vaccination can exhibit \emph{hysteresis}~\cite{chen2019imperfect}, namely, the dependence of population vaccination rate on its past trajectory. Such hysteresis effect makes the population sensitive to changes in factors that drives vaccination behavior, such as cost and effectiveness. The presence of hysteresis also can hinder the recovery of vaccine uptake, in spite of decrease in the perceived cost of vaccination or improvement of vaccine efficacy, as the vaccination trajectory can get stuck in the hysteresis loop.

Among others, over the past decade, researchers have proposed \emph{behavioral epidemiology} as a means of integrating the study of epidemiology with the influence of human behavior including but limited to health decisions made by individual actors responding to infection risks~\cite{bauch2013behavioral,galvani2016human}. As such, a feedback loop exists between health behaviors and the spread of an epidemic: individuals may take preventative measures, such as vaccination or reduced contact with others, in response to perceived risks. These responses in turn modify the spread of infection. The interplay between changing opinions of vaccination and epidemic spreading on social networks constitutes a ``dueling contagion'' process~\cite{fu2017dueling,chen2022highly,fugenschuh2023overcoming}. It is of fundamental significance to achieve a comprehensive understanding of the rich dynamics generated by this sort of dueling contagion~\cite{bauch2013social,glaubitz2024social}.

In particular, the use of behavior-disease interaction models has become an important approach to study how vaccine compliance can be influenced by a wide range of factors~\cite{bauch2005imitation,reluga2006evolving,vardavas2007can,wu2011imperfect, fu2011imitation,zhang2012rational,ndeffo2012impact,shim2012influence,cardillo2013evolutionary,wu2013peer,oraby2014influence,zhang2014effects,wang2016statistical,chen2019imperfect,khan2023time,de2024interplay,lu2023reinforcement,glaubitz2023population,he2024effect}, ranging from vaccine scares~\cite{bauch2012evolutionary} to disease awareness~\cite{wang2016suppressing}. Prior work shows that a misalignment between individual interest and the population interest can cause suboptimal vaccination coverage~\cite{fine1986individual,bauch2003group,bauch2004vaccination,galvani2007long,cornforth2011erratic}, thereby leading to a tragedy of the commons in vaccination uptake~\cite{hardin1968tragedy,shen2023committed}.

An individual's vaccination contributes to herd immunity, meaning these who forgo vaccination can be indirectly protected by the presence of herd immunity. The problem of vaccine compliance is thus often represented as a public-goods dilemma. A misalignment between individual self-interest and population interest can yield the ``free rider'' problem in vaccine uptake~\cite{perisic2009social,breban2007mean,fine1986individual,bauch2003group,basu2008integrating,larson2014understanding}, thereby causing suboptimal vaccination coverage~\cite{omer2009vaccine}. Moreover, the long-standing dilemma of voluntary vaccination is exacerbated by spreading concerns about vaccine safety and efficacy~\cite{chen1999vaccine,amanna2005public,hughes2006news,coelho2009dynamic,wu2011imperfect,chen2019imperfect,hu2024evolutionary}. Recently, considerable attention has been paid to improving our understanding of the role of social factors in epidemiology~\cite{salathe2008effect,funk2010modelling,bauch2013social,wang2016statistical,fu2017dueling,chen2022highly,funk2009spread,eames2009networks,saad2023dynamics,glaubitz2024social,shi2024determinants,espinoza2024adaptive,saad2024impact}.

Understanding the impact of social networks on public health behavior and especially vaccination choices is of particular interest~\cite{bauch2013social}. A vaccine's success can become its own demise. Once the incidence of vaccine-preventable common childhood diseases becomes rare, parents who are unfamiliar with the diseases pay more attention to concerns regarding the risks of vaccination rather than the disease itself~\cite{wadman2017vaccine}. This leads to social contagion of vaccine scares or skepticisms, which can hinder vaccination efforts~\cite{salathe2008effect,salathe2011assessing}. Vaccine-averse attitudes amplify the costs of vaccination due to heightened concerns about vaccine safety and risks~\cite{hughes2006news}. The issue is further complicated by anecdotes or personal experience regarding vaccine effectiveness, as vaccines may be imperfect and confers only partial protection against diseases~\cite{chen2019imperfect,gandon2001imperfect}. The impact of heterogeneous beliefs on vaccination dynamics remains poorly understood, particularly when a small fraction of individuals hold pre-existing, fixed belief as opposed to vaccine-neutral attitudes.  

Motivated by these considerations, here we investigate the social imitation dynamics of vaccination in well-mixed and spatial populations driven by vaccine effectiveness and beliefs. We quantify and compare the sensitivity and fragility of vaccination coverage in the presence of pre-existing or coevolving vaccine beliefs. Our work shows that even a small fraction of individuals with fixed vaccine-averse beliefs can exacerbate the hysteresis effect, causing vaccination coverage to be more sensitive to changes in the perceived cost of vaccination and vaccine effectiveness, compared to cases without any beliefs. Furthermore, the coevolution of vaccine beliefs and behavior choices has a lesser impact than rigid, fixed beliefs but still reduces the population's resilience to perturbations in perceived vaccination cost and effectiveness. By revealing the interference between epidemic spreading and the social contagion process of vaccine beliefs that shape public perceptions of vaccine safety and risk, our work provides deep insights into the social factors that drive vaccination decisions and as well as barriers to boosting vaccine uptake.

\begin{figure}[!h]
\includegraphics[width=\linewidth]{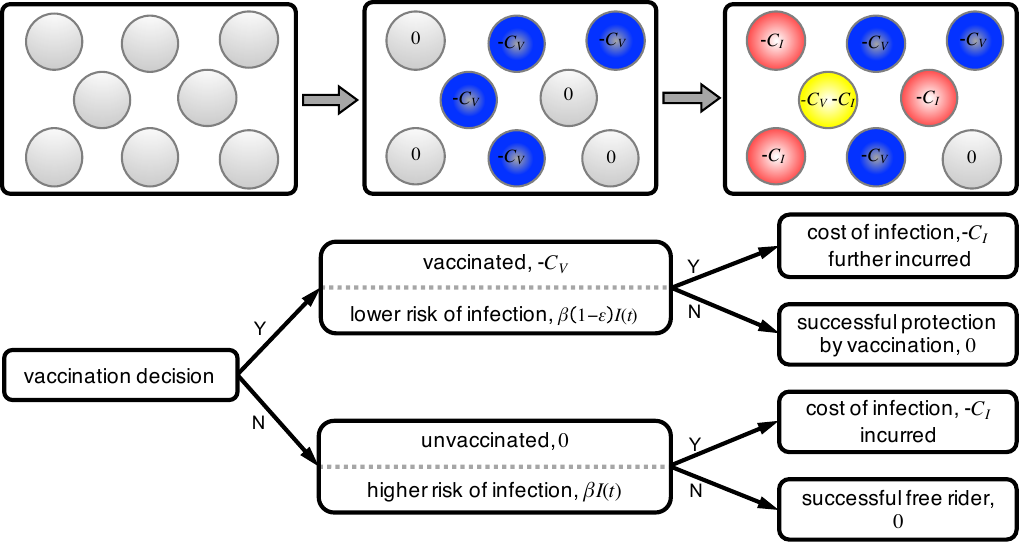}
\caption{{\bf Model schematic.}
We model the vaccination dynamics under imperfect vaccine as a two-stage game. At stage 1 (vaccination choice), a proportion $V_0$ of the population decides to vaccinate. Vaccination costs $C_v$ and provides \emph{imperfect} protection against the infectious disease. At stage 2 (health outcome), we use the Susceptible-Infected-Recovered (SIR) model with preemptive vaccination to simulate the epidemiological process. Every individual faces the risk of infection, which depends on their vaccination status and the number of infectious neighbors, $I(t)$, they have. The transmission rate of the disease (per day per infectious neighbor) to unvaccinated individuals is $\beta$, as compared to $\beta (1 - \varepsilon)$ for vaccinated. Here, the parameter $\varepsilon \in [0, 1]$ denotes the level of vaccine effectiveness. The cost of infection is $C_I$. Without loss of generality, we use the relative cost of vaccination, $c = C_v/C_I\in [0, 1]$ in the remainder of this paper. Those unvaccinated individuals who remain healthy are free-riders off the vaccination efforts of others, as they are indirectly protected to some extent by herd immunity. }
\label{fig1}
\end{figure}

\section*{Results}

To begin, we first study how social imitation dynamics of vaccination, where individuals are not perfectly rational, can be impacted by the presence of an imperfect vaccine. In addition to weighing the perceived cost of vaccination with the risk of infection, the effectiveness of vaccination is also an important factor driving vaccination decisions. Specifically, we model the vaccination dynamics under imperfect vaccines as a two-stage game as shown in Fig.~\ref{fig1}. At stage 1 (vaccination choice), a proportion $V_0$ of the population decides to vaccinate. Vaccination costs $C_v$ and provides \emph{imperfect} protection against the infectious disease. At stage 2 (health outcome), we use the Susceptible-Infected-Recovered (SIR) model with preemptive vaccination in stage 1 to simulate the epidemiological process:
\begin{equation}
\begin{aligned}
\frac{dS}{dt} & =  - \beta S I,\\
\frac{dI}{dt} & = \beta SI + (1-\varepsilon)\beta VI - \gamma I,\\
\frac{dV}{dt} & =  -(1-\varepsilon)\beta V I,\\
\frac{dR}{dt} & = \gamma I.
\end{aligned}
\label{simplesir}
\end{equation}
Here, the transmission rate of the disease to an unvaccinated individual is $\beta$, as compared to $(1-\varepsilon)\beta$ for a vaccinated individual. Denote by $\gamma$ the disease recovery rate. Every individual faces the risk of infection, which depends on their vaccination status and the number of infectious neighbors, $I(t)$, they have. Thus the parameter $\varepsilon$ quantifies the effectiveness of vaccination in protecting against the disease. For perfect vaccines, $\varepsilon = 1$, vaccinated individuals have zero risk of infection, as already analyzed in previous studies~\cite{fu2011imitation}. For imperfect vaccines, $0< \varepsilon < 1$, vaccinated individuals still face the risk of getting infected but with a reduced likelihood compared to unvaccinated individuals~\cite{chen2019imperfect}. The cost of infection is $C_I$. Those unvaccinated individuals who remain healthy are free-riders off the vaccination efforts of others, as they are indirectly protected to some extent by herd immunity~\cite{fu2011imitation}. 

Without loss of generality, we assume the relative cost of vaccination to infection is $c = C_v/C_I \in (0, 1)$. As shown in  Fig.~\ref{fig1}, there are four possible health outcomes, and thus, we have four different payoffs at the end of the current season. Up to a positive constant factor, the payoff for a vaccinated individual who remained healthy during the epidemic (denoted by the fraction $x_0$) is $-c$, the payoff for a vaccinated individual who still contracted the disease (denoted by the fraction $x_1$) is $-1-c$; the payoff for an unvaccinated individual who became infected (denoted by the fraction $y_1$) is $-1$ while the payoff for an unvaccinated individual who remained healthy (denoted by the fraction $y_0$) is $0$. The proportions of these four types of outcomes are determined by the SIR-V dynamics as given in Eq.~\eqref{simplesir} and the vaccination level $x$ (see Materials and Methods).  In stage 2, individuals revisit their vaccination choices through social imitation, a social learning process based on pairwise comparison~\cite{traulsen2007pairwise}. An individual $i$ with strategy $S_i$ and payoff $\pi_i$ randomly chooses one of their neighbors $j$ with strategy $S_j$ and payoff $\pi_j$ and switches to $j$'s vaccination choice with the probability $\phi_{S_j \to S_i}$ given by the Fermi function:
\begin{equation}
\phi_{S_j \to S_i} = \frac{1}{1+ \exp[-K(\pi_j - \pi_i)]},
\end{equation}
where the parameter $K > 0$ is the intensity of selection and quantifies the level of rationality. In a well-mixed population, the replicator-like equation for the evolution of vaccination choice is given by:
\begin{align}
\dot{x} & = x_0 y_0 \tanh[\frac{K}{2}(-c - 0)] + x_0 y_1\tanh[\frac{K}{2}(-c + 1)] \\
& + x_1 y_0 \tanh[\frac{K}{2}(-c-1 - 0)] + x_1 y_1\tanh[\frac{K}{2}(-c-1 +1)], 
\end{align}
where $x_0 + x_1 = x$ and $y_0 + y_1 = 1 -x$.

\begin{figure}[!h]
\includegraphics[width=\linewidth]{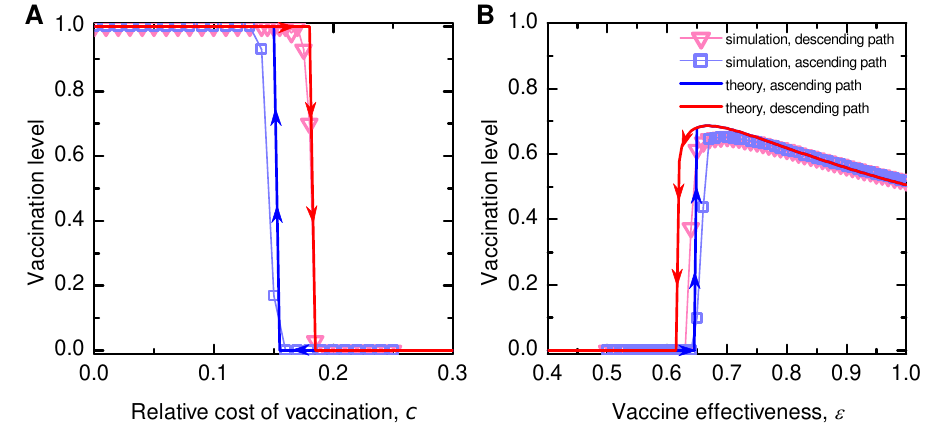}
\caption{{\bf Bistability of equilibrium vaccination levels and the emergence of hysteresis loops in well-mixed populations.} We simulate the social imitation of vaccination dynamics in a finite, well-mixed population and obtain hysteresis loops, composed of the ascending and descending paths, in response to variations to model parameters: (\textbf{A}) the relative cost of vaccination, $c$, and (\textbf{B}) vaccine effectiveness, $\varepsilon$. The population can exhibit bistability within certain ranges of the model parameters $c$ and $\varepsilon$. Stochastic agent-based simulation results align with theoretical analysis, with noticeable discrepancies attributable to finite population effects. Parameters: population size $N = 1000$, infection seeds $I_0 = 10$. For the first simulation, the initial number of vaccinated individuals is $V_0 = 500$. For subsequent simulations, the results of the preceding simulations are used as the initial conditions, while the model parameters are varied in a prescribed sequence of increasing and then decreasing values. Disease transmission rate $\beta N = 0.25$, recovery rate $\gamma = 0.1$,  intensity of selection $K = 1$. (A) effectiveness $\varepsilon = 0.4$, (B) $c = 0.35$. Simulation results are averaged over 100 independent runs.
} 
\label{fig2}
\end{figure}

We now proceed with bifurcation analysis and identify parameter regions that enable bistability and thus allow the hysteresis loop to occur. We present stochastic agent-based simulation results in Fig.~\ref{fig2} along with the numerical theoretical analysis based on the system of differential equations. As predicted by the theoretical analysis, the population exhibits bistability with respect to varying the relative cost of vaccination $c$ and vaccination effectiveness $\varepsilon$ and the population abruptly transitions from high vaccination level to complete opt-outs when $c$ increases beyond a threshold and $\varepsilon$ drops below a threshold (the descending path). However, to recover the vaccination level to previously high coverage, the ascending path takes a different route instead of reversing the descending path, requiring an even lower threshold in cost and higher vaccine effectiveness. Here, we confirm the occurrence of this hysteresis effect for social learning that allows irrationality (with $K = 1$ in Fig.~\ref{fig2}), where individuals imitate with higher probability those with higher payoffs but still can imitate those with lower payoffs.

Aside from well-mixed populations, we also study vaccination dynamics in spatial populations, where, for example, individuals are situated on a square lattice with the von Neumann neighborhood~\cite{szabo1998evolutionary}. Such population structure restricts whom individuals can imitate, or be infected by, to just their immediate neighbors. Individuals' vaccination decisions and health outcomes determine their payoffs. They can revisit their vaccination choices by imitating more successful strategies among their immediate neighbors. It is straightforward to study vaccination dynamics in a variety of networks, including random networks and scale-free networks~\cite{fu2011imitation}. In general, we confirm the existence of a threshold for vaccine effectiveness $\varepsilon$, below which multiple stable vaccination equilibria emerge. We use simulations to determine whether population structure, as compared to the well-mixed case, can strengthen the \emph{hysteresis effect}, thereby further hindering the recovery of vaccine uptake. In what follows, we detail these results to better understand the role of population structure in vaccine uptake behavior, especially in the presence of imperfect vaccines as well as fixed or coevolving vaccine beliefs. 

\begin{figure}[!h]
\includegraphics[width=\linewidth]{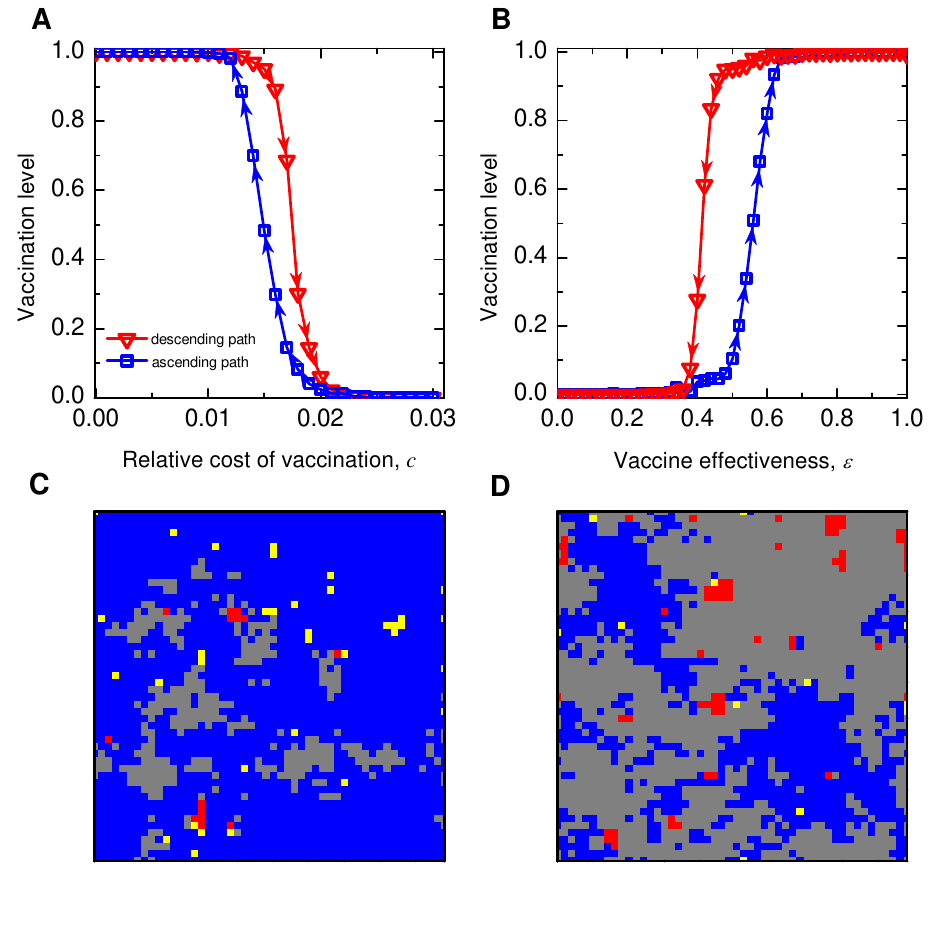}
\caption{{\bf Hysteresis and sensitivity of vaccination dynamics in lattice populations.} (A) and (B) depict the hysteresis loops, represented by the descending and ascending paths of the population's equilibrium vaccination level, in response to variations in the relative cost of vaccination, $c$, and vaccine effectiveness, $\varepsilon$, respectively. Overall, the critical parameter region where the hysteresis occurs depends on specific model parameter choices. The spatial population is highly sensitive to changes in $c$ and $\varepsilon$ and abruptly transitions between complete opt-out and universal coverage of vaccination. (C) and (D) show spatial snapshots of population states along the descending and ascending paths of Fig.~\ref{fig3}A for a fixed $c = 0.015$, respectively. The color codes of individuals are the same as in Fig.~\ref{fig1}: Blue represents individuals who were vaccinated and remained healthy during the season,
yellow represents individuals who were vaccinated but still became infected, grey represents individuals who were unvaccinated yet remained healthy, and red represents individuals who were unvaccinated and became infected. Parameters:  Square lattice $N = 50\times 50$ with the von Neumann neighborhood, infection seeds $I_0 = 30$, initial number of vaccinated $V_0 = 1250$, disease transmission rate $\beta = 0.04$, recovery rate $\gamma = 0.1$, intensity of selection $K = 1$. (A) effectiveness $\varepsilon = 0.8$ (B) $c = 0.01$. (C) (D): $c = 0.015$, $\varepsilon = 0.8$. Simulation results are averaged over 150 independent runs.
}
\label{fig3}
\end{figure}

We observe a high sensitivity of vaccination coverage in structured populations. For comparable disease impact without vaccination, the spatial population manifests drastic fragility to changes in the perceived cost of vaccination, even for more effective vaccination (Fig.~\ref{fig3}A) and requires much more improvement in vaccination effectiveness in order to boost high vaccination levels, even for a smaller cost of vaccination (Fig.~\ref{fig3}B). One reason is that population structure promotes assortment, and clusters of unvaccinated and vaccinated individuals together (see Fig.~\ref{fig3}C and D) strengthen the vaccination coverage's sensitivity to changes, indicating a double-edged sword effect of population structure~\cite{fu2011imitation}.

We emphasize that social contagions may also promote the spread of misinformation and bad health behaviors, as well~\cite{salathe2011assessing}.  The spread of vaccine scares among parents (via social contagion) has caused the vaccination rates of newborns to plunge from high levels~\cite{hughes2006news}, which, in turn, has increased the incidence of several childhood diseases (via biological contagion). These fears are fueled not only by face-to-face interaction, but also by changing opinions of vaccination that are expressed in online social media~\cite{salathe2011assessing}. It is critical for us to better understand these spreading processes so public health efforts can take advantage of the positive effects of social contagion, while ameliorating its potential negative impacts.

\begin{figure}[!h]
\includegraphics[width=\linewidth]{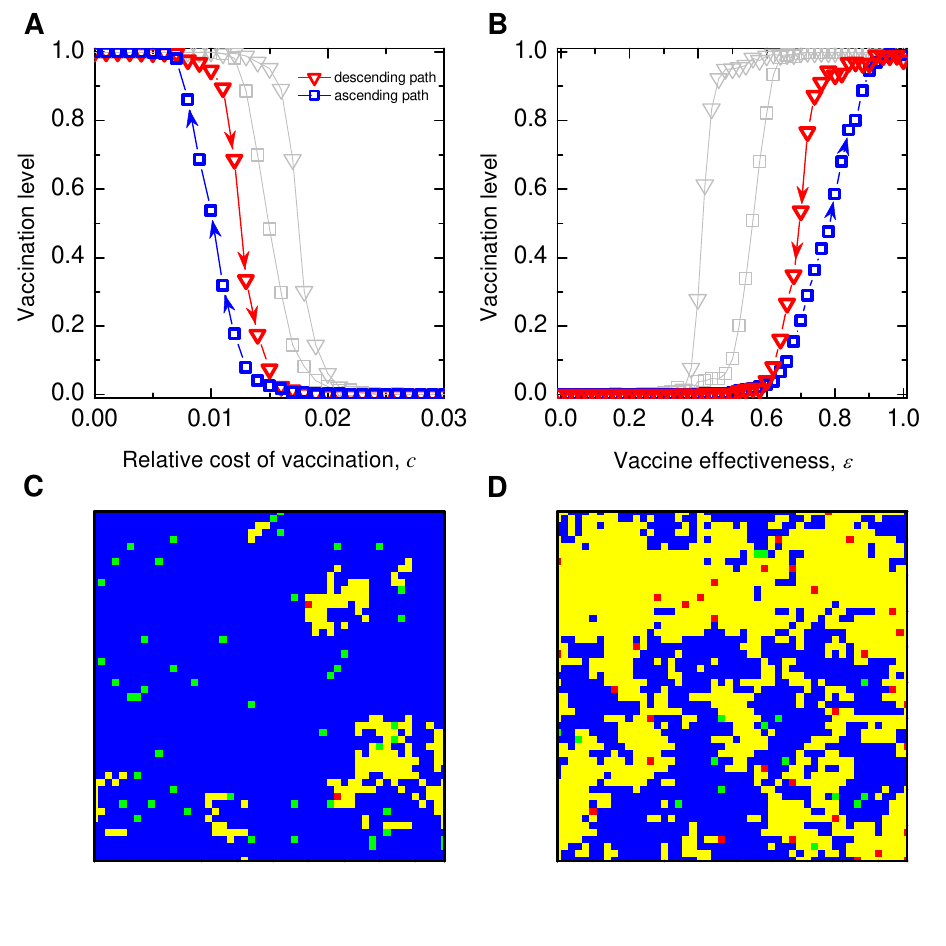}
\caption{{\bf Impact of pre-existing, fixed vaccine beliefs on vaccination dynamics.} The presence of vaccine-averse (skeptical) beliefs, even at low frequencies, can render the population more sensitive to the cost of vaccination and vaccine effectiveness. Shown are the hysteresis loops of vaccination levels with respect to changes in (A) the cost of vaccination and (B) the vaccine effectiveness. For comparison sake, the grey lines are the results in Fig.~\ref{fig2} without any vaccine beliefs. (C) and (D) show spatial snapshots of population states in the descending and ascending paths respectively. The color of individuals indicates their specific combinations of vaccine beliefs and uptake behaviors: blue: vaccinated individuals with vaccine-neutral attitude; yellow: unvaccinated individuals with vaccine-neutral attitude; green: vaccinated individuals with vaccine-averse attitude; red: unvaccinated individuals with vaccine-averse attitude. Parameters: square lattice $50\times 50$ with von Neumann neighborhood, initial number of infection seeds $I_0 = 30$, initial number of vaccinated $V_0 = 1250$, fixed number of vaccine skeptical individuals $50$, disease transmission rate $\beta = 0.04$, recovery rate $\gamma = 0.1$, $\theta = 0.1$, $K = 1$. (A) effectiveness $\varepsilon = 0.8$ (B) $c = 0.01$, (C) (D): $c = 0.01$, $\varepsilon = 0.8$. Simulation results are averaged over 150 independent runs.
}
\label{fig4}
\end{figure}

We first study the scenario when a fixed population of individuals believes vaccines impose elevated risks. To inform our choice of the proportion of negative vaccine views, we use the recent Pew Research data, which shows that 7\% of respondents believe the preventive health benefits of MMR vaccines are low~\cite{PewVac}. To incorporate this consideration, we introduce an additional perceived vaccination cost, $\theta$, for this subpopulation in our model.
To illustrate how a small proportion of individuals with vaccine-averse beliefs can disproportionately impact vaccine uptake ,  we set their proportion as low as 2\% in our simulations. 
Despite such a small fraction of individuals with fixed beliefs about the amplified cost of vaccination (with $\theta = 0.1$), the population exhibits much higher sensitivity and fragility of high vaccination levels to perturbations. Fig.~\ref{fig4} demonstrates that the occurrence of the hysteresis loop at a much smaller cost of vaccination (Fig.~\ref{fig4}A) and at a much higher vaccination effectiveness (Fig.~\ref{fig4}B), reducing the critical thresholds of relevant vaccine parameters by more than half (cf. Fig.~\ref{fig3}). 

The corresponding spatial snapshots in Fig.~\ref{fig4}C and D reveal that the opt-out behavior of vaccine-skeptical individuals can spread to their neighbors, leading to noncompliance even among vaccine-neutral individuals. Conversely, vaccination by those with neutral attitudes can convert those individuals holding vaccine-averse views. Taken together, the presence of rigid fixed beliefs, even in a tiny proportion of the population, can render the population significantly more sensitive to changes compared to cases without such beliefs.

\begin{figure}[!h]
\includegraphics[width=\linewidth]{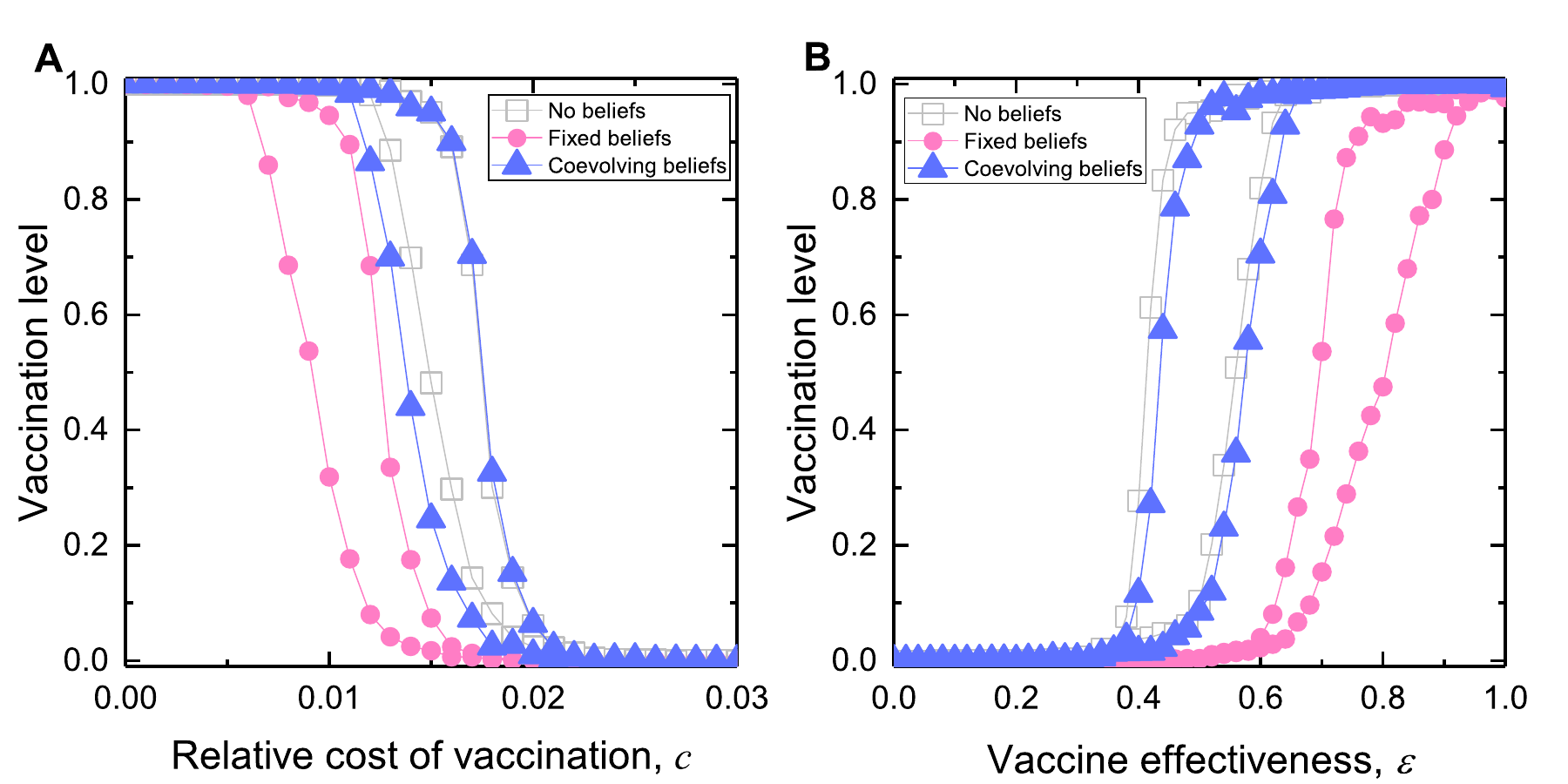}
\caption{{\bf Coevolution of vaccine beliefs and uptake behavior.} Plotted are the hysteresis loops of equilibrium vaccination levels as a function of (\textbf{A}) the relative cost of vaccination, $c$, and (\textbf{B}) vaccine effectiveness, $\varepsilon$. For comparison, we include the simulation results in the absence of any vaccine beliefs (Fig.~\ref{fig3}) as well as those with fixed vaccine beliefs (Fig.~\ref{fig4}). Compared to the case without vaccine beliefs, the concurrent spreading of beliefs--where a vaccine-neutral attitude competes with a vaccine-averse attitude alongside the social contagion (imitation) process of vaccine behavior choices--lead to slightly less favorable condition for vaccination. However, this impact is much less severe than in the scenario where a small proportion of individuals hold a vaccine-averse attitude and remain unchanged. Parameters: (\textbf{A}, \textbf{B}) square lattice $50\times 50$ with von Neumann neighborhood, initial number of infection seeds $I_0 = 30$, initial number of vaccinated $V_0 = 1250$ (50\%), initial number of vaccine skeptical individuals $1250$ (50\%), disease transmission rate $\beta = 0.04$, recovery rate $\gamma = 0.1$, $\theta = 0.1$, $K = 1$. (A) for fixed effectiveness $\varepsilon = 0.8$, and (B) for fixed $c = 0.01$.
Simulation results are averaged over 150 independent runs.
}
\label{fig5}
\end{figure}

Finally we study the scenario where competing beliefs spread interpersonally and coevolve with vaccination behavior in a manner similar to disease transmission (Fig.~\ref{fig5}). Individuals can revisit and change both their vaccine beliefs and behavior choices via social contagion. The combination of beliefs and behavior results in coevolutionary dynamics of four types, coupled with disease spreading. Just as before, a particular combination of vaccine belief and behavior is more likely to spread whenever it yields higher payoffs. Unlike fixed beliefs, this extended scenario allows individuals to adjust their vaccine beliefs in addition to their vaccination decisions, based on their own experiences and peer influence.  The coevolution of belief and behavior may lead to belief-behavior consistency through self-correcting social imitation. That said, while the concurrent spreading of beliefs and behavior choices results in slightly less favorable conditions for the stability of vaccination coverage, it still fares much better than the case with fixed beliefs (see Fig.~\ref{fig5}).

\begin{figure}[!h]
\includegraphics[width=\linewidth]{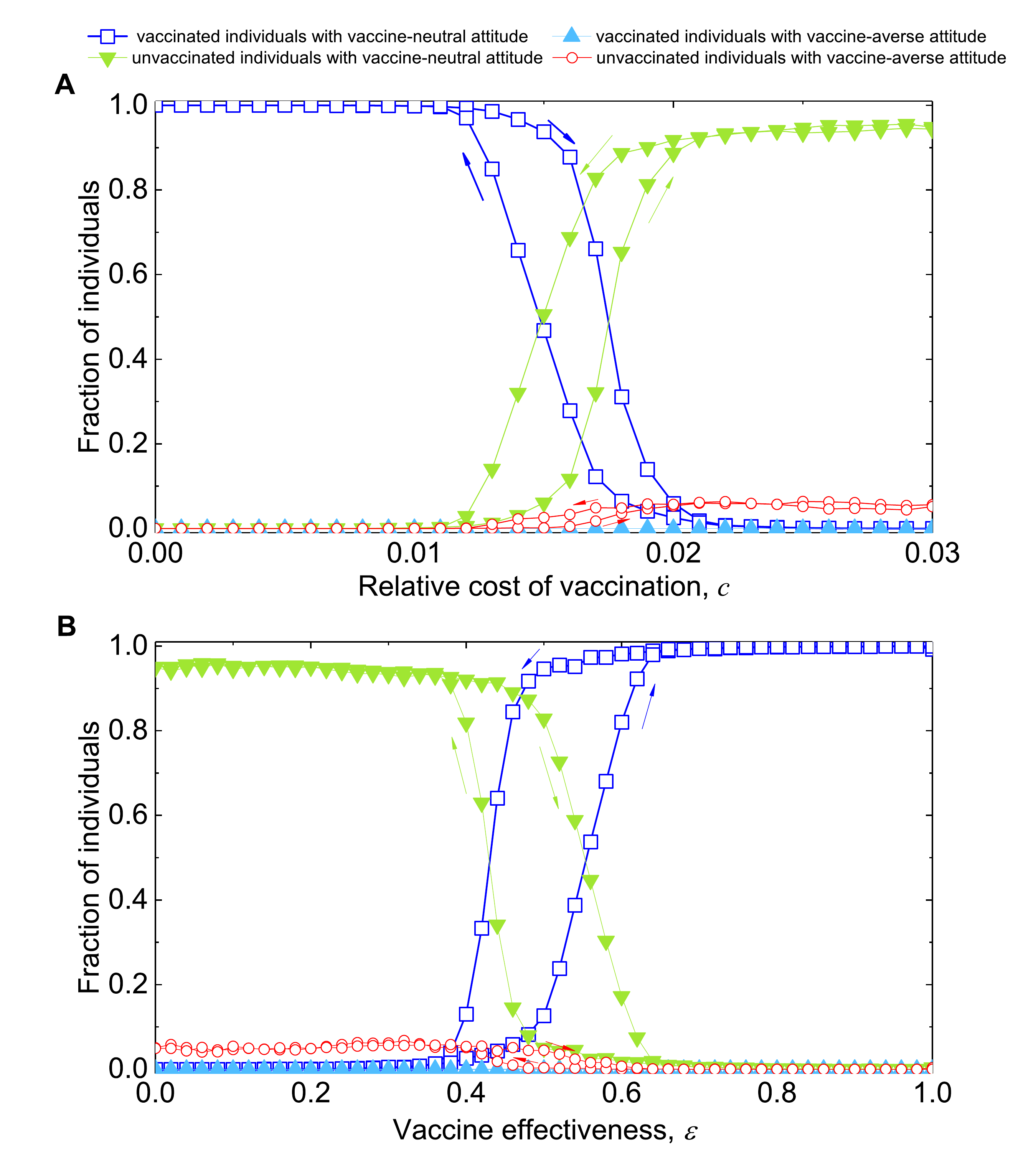}
\caption{{\bf Microscopic view of the hysteresis loops arising from the coevolution of vaccine beliefs and uptake behavior.} Shown are hysteresis loops, indicated by the corresponding descending and ascending paths of the equilibrium fractions of individuals grouped into four types based on the combinations of their vaccine beliefs and behavior choices, as a function of (\textbf{A}) the relative vaccination cost $c$ and (\textbf{B}) the vaccine effectiveness $\varepsilon$. The equilibrium fraction of individuals with a vaccine-averse attitude who also opt for vaccination is almost zero across the parameter space studied. The proportion of vaccine-averse individuals almost exclusively opt out of vaccination, and their fraction can reach a maximum of around 6\% in the population. Parameters: square lattice $50\times 50$ with von Neumann neighborhood, initial number of infection seeds $I_0 = 30$, initial number of vaccinated $V_0 = 1250$ (50\%), initial number of vaccine skeptical individuals $1250$ (50\%), disease transmission rate $\beta = 0.04$, recovery rate $\gamma = 0.1$, $\theta = 0.1$, $K = 1$. (\textbf{A}) for fixed effectiveness $\varepsilon = 0.8$ (\textbf{B}) for fixed $c = 0.01$.
Simulation results are averaged over 150 independent runs.
}
\label{fig6}
\end{figure}

Fig.~\ref{fig6} provides a detailed view of the four types of individuals via their hysteresis loops, rather than their overall vaccination levels in Fig.~\ref{fig5}. It is notable that unvaccinated individuals with vaccine-averse attitudes arise in the population, whereas individuals with vaccine-neutral attitudes seldom opt for vaccination. The spread of vaccine-averse beliefs can reach a maximum at 6\% under extremely unfavorable conditions for vaccination, characterized by high perceived costs $c$ and low vaccine effectiveness $\varepsilon$. Vaccine-averse individuals are almost exclusively noncompliant with vaccination. Fig.~\ref{fig6} also suggests that the emergence of vaccine-averse attitudes arises and persists in the population, driven by the perceived cost of vaccination and compromised vaccine effectiveness. The descending path to mitigate these opinions requires much higher levels of improvement in vaccine confidence and trust. These results highlight that it is crucial for public health to harness the power of social contagion -- which, while exhibiting a dual impact -- to ameliorate the impact of vaccine-averse beliefs and boost vaccination confidence and demand.  

\section*{Discussion and Conclusion}

Compared to non-pharmaceutical interventions requiring repeated adherence, vaccination is typically one-off action that can provide sufficient protection  during the ongoing epidemic~\cite{bauch2013behavioral}. Under noncompulsory, voluntary vaccination,  populations can achieve high vaccination coverage for small costs of vaccination, provided the vaccine is effective. Complications arise when exaggerated perceived risk or cost of vaccination---due to adverse side effect though uncommon---and concerns about vaccine safety undermine public confidence~\cite{wadman2017vaccine}. Such beliefs erode individuals' intention to vaccinate~\cite{coelho2009dynamic,gowda2013rise}. In this work, we study how the presence of small fraction of vaccine-averse beliefs can disproportionately impact  vaccination coverage's sensitivity and fragility to small changes in perceived vaccination cost and vaccine effectiveness. 

It is not surprising that the spread of such beliefs occurs alongside individual vaccination behavior~\cite{salathe2008effect,salathe2011assessing}. Vaccine-averse beliefs can gain a foothold in the population, especially when the perceived cost of vaccination increases or vaccine effectiveness is comprised~\cite{hughes2006news}. Beyond the hysteresis loop of vaccination levels, our work demonstrates the existence of a similar hysteresis effect for vaccine beliefs, presenting an additional roadblock to efforts to increase vaccination rates. Individuals are not perfectly rational, and skeptical beliefs about vaccines can replace fear of disease. Ironically, because vaccines reduce the incidence and severity of diseases, people become overly focused on anecdotes and occasional vaccination failures~\cite{amanna2005public}. Therefore, effectively communicating the efficacy and cost-effectiveness of vaccination become pivotal ---not to intentionally create pro-vaccine beliefs but at least to foster vaccine-neutral attitudes, as our results suggest. 

In this study, we examine the interplay between individual beliefs, vaccination decisions, and social and biological contagion processes. Our model incorporates boundedly rational decision-making driven by social imitation with incomplete information, capturing the nuances of human behavior influenced by vaccine-related beliefs. We analyze how the presence, as well as concurrent spread, of vaccine skepticism in spatial populations can amplify small increases in perceived vaccination costs, leading to significant declines in vaccine uptake. By coupling belief-behavior coevolution with epidemiological dynamics, our model highlights the indirect pathways through which beliefs (misinformation) about vaccines propagates and undermines public health efforts.

Our findings highlights the importance of addressing vaccine hesitancy as both a behavioral and social phenomenon~\cite{bauch2013social}. Through simulations, we demonstrate that even a modest fraction of vaccine-averse individuals can create cascading effects that significantly impact the fragility of herd immunity thresholds. Moreover, we identify potential targeted network interventions--such as effective communication campaigns that reduce the fear of vaccination or incentives that subsidize the cost of vaccination---can mitigate the spread of vaccine skepticism. These insights provide a pathway for designing more resilient public health strategies that address both the direct epidemiological challenges of vaccine-preventable diseases and the social dimensions of belief formation~\cite{fugenschuh2023overcoming}.

The resurgence of vaccine-preventable disease outbreaks underscores the tradeoff between individual choices and collective responsibility~\cite{fine1986individual}. Fundamentally, the vaccination dilemma is one example of real-world human cooperation problems, among many others~\cite{wang2020eco}. We are interconnected, and so is our health~\cite{smith2008social}. Leveraging the power of social contagion to overcome vaccine hesitancy is key to avoiding the tragedy of the commons in biological contagions. Our work provides model-based insights into the presence and spread of vaccine related beliefs in social networks and their impact on health interventions aimed at improving vaccine compliance.

\section*{Materials and Methods}

In view of the recurrent outbreaks of infections such as influenza and vaccination effectiveness, we use an evolutionary game-theoretical approach to study the seasonal vaccination game. A feedback loop exists between the vaccination decisions of individuals and their health outcomes. Disease incidence is influenced by vaccination behavior: a high level of vaccine coverage can reduce disease incidence to very low levels, which in turn lowers the perceived risk of infection and reduces the demand for vaccination~\cite{bauch2013behavioral}. As vaccination coverage drops, the number of susceptible individuals increases. When the proportion of susceptible, unvaccinated individuals exceeds a threshold (the complement of the herd immunity threshold), an outbreak of infectious disease can occur~\cite{anderson1991infectious}. A surge in disease incidence can, in turn, convince individuals to start vaccinating again~\cite{reluga2006evolving}. 

The vaccination dilemma game consists of two stages: a vaccination decision-making process at the beginning of the season, followed by the disease epidemic \cite{fu2011imitation,chen2019imperfect}. The proposed model is illustrated in Fig~\ref{fig1}. During the first stage, individuals make a preemptive vaccination decision, choosing whether or not to get vaccinated based on social imitation of peers' choices, taking into account the costs of vaccination and infection. A vaccinated individual pays a cost $C_v>0$ while an unvaccinated individual incurs no direct upfront  cost. This vaccination cost includes the time spent in receiving the vaccine, the perceived risks of vaccination, long-term health impacts, and other intangible factors. Denote by $x$ the vaccination level at the end of stage 1. 

During the epidemic season (stage 2), the epidemic is initiated by a number $I_0$ of infected individuals and then spreads in the population (both well-mixed and lattice populations) according to the susceptible-infected-recovered-vaccinated (SIR-V) dynamics, with a per day per infected neighbor transmission rate $\beta$ and a per day recovery rate $\gamma$. The basic reproduction number \( R_0 \) is defined as $R_0 = \beta/\gamma$.
Let vaccination effectiveness be $\varepsilon$; then the vaccinated population has a reduced transmission rate of $\beta(1-\varepsilon)$. The epidemic continues until there are no more newly infected individuals (which typically occurs in under 200 days for all cases simulated). The SIR-V epidemiological process is simulated by the Gillespie algorithm~\cite{fu2011imitation}. Once the epidemic ends, individuals can revisit their vaccination decisions for the next season. 

The final epidemic size $R(\infty)$ satisfies the transcendental equation:
\begin{equation}
x \left( \exp(-R_0 R(\infty)) - \exp(-(1 - \varepsilon) R_0 R(\infty)) \right) - R(\infty) + 1 - \exp(-R_0 R(\infty)) = 0.
\end{equation}
At the end of the season, the relative fraction of individuals infected ($x_1$) among vaccinated individuals ($x$) is 
\begin{equation}
\frac{x_1}{x} = 1 - \exp\left(-(1 - \varepsilon) R_0 R(\infty) \right),
\end{equation}
while the relative fraction of individuals infected ($y_1$) among unvaccinated individuals ($y$) is
\begin{equation}
\frac{y_1}{y} = 1 - \exp\left(-R_0 R(\infty) \right).
\end{equation}

Infection incurs a cost $C_I>0$, which includes expenses and time for health care as well as an elevated chance of mortality. Without loss of generality, we use the relative cost of vaccination $c  = C_v/C_I$, while setting $C_I = 1$. Thus, $0 \leq c = C_v \leq 1$. 

Individuals adjust their vaccination strategies by imitation, where successful individual's strategy is more likely to be followed~\cite{bauch2005imitation,fu2011imitation,ndeffo2012impact,bauch2012evolutionary}. An individual's imitation behavior is based on the current payoff difference between herself and a randomly selected neighbor. If the strategy of the selected neighbor has a higher payoff than her own strategy in the past epidemic season, then the individual imitates her neighbor's strategy with a higher probability. In this work, we use the Fermi function to determine the probability of imitation, accounting for bounded rationality in the decision process~\cite{blume1993statistical,szabo1998evolutionary,traulsen2010human}. Specifically, Individual $i$ randomly selects one neighbor $j$ from her immediate neighborhood. The probability that individual $i$ adopts individual $j$'s strategy is given by \cite{blume1993statistical,szabo1998evolutionary,traulsen2010human}:

\begin{eqnarray}
\label{eq:1}
W(S_j \to S_i)=\frac{1}{1+\exp(-K(\pi_j - \pi_i))},
\end{eqnarray}
where $S_i$ means the vaccination choice for individual $i$: vaccination or non-vaccination. $\pi_i$ denotes the current payoff of individual $i$ at the end of current season. For $i$'s payoff, we have four possible outcomes:
\begin{center}
\begin{itemize}
	\item $\pi_i = - c$ if  $i$ is vaccinated and is not infected;
	\item $\pi_i = - c - 1$ if $i$ is vaccinated and is infected;
	\item $\pi_i = -1$ if $i$ is not vaccinated and is infected;
	\item $\pi_i = 0$ if $i$ is not vaccinated and is not infected.
\end{itemize}
\end{center}
The parameter $K$ is the intensity of selection, indicating how strongly individuals are responsive to payoff difference. This updating rule diverges from a perfect rationality model. Here we adopt $K=1$ for our simulations~\cite{fu2011imitation}. It is worth noting that individuals adjust their strategies based on the realized payoffs, not expected payoffs. In a population with low vaccination uptake, many non-vaccinators fall ill, but if individual $i$ happens to choose one of the few successful free riders as a role model, then she will be more likely to imitate the free rider's strategy~\cite{fu2011imitation}. 

To obtain hysteresis loops with respect to varying relevant model parameters for our base model, we run a sequence of simulations with the model parameter varied in both increasing and decreasing order. For the first simulation in each sequence (corresponding to ascending or descending path of the identified hysteresis loop), we use the same initial state, which consists of a fraction $V_0$ of vaccinated individuals, randomly distributed throughout the population. Each two-stage iteration (vaccination strategy updating followed by an epidemic process) updates the proportion of vaccination strategies. For the stochastic epidemiological process, where the number of initial infection seeds is denoted as $I_0$, we use the Gillespie algorithm with a maximum duration of $10^4$ to ensure the complete exhaustion of every infected individual. The equilibrium results are obtained by averaging over the last 1000 iterations from a total of 4000 iterations, and each data point reported in this paper is the result of an average of at least 100 independent realizations. For our extended models with vaccine beliefs, we use similar simulation procedures with the model parameters and initial conditions specified in the respective figure captions.

In our base model, we assume that individuals have homogenous perceptions of the cost of vaccination: $c$ is the same for all individuals. In our extended model, we consider two groups of individuals holding vaccine-neutral versus vaccine-averse attitudes. For the latter, an additional vaccination cost $\theta>0$ is incurred. The perceived payoffs differ for these two groups as follows. For vaccine-neutral individuals, the payoffs are the same as those described above. For vaccine-averse individuals, we have:
\begin{center}
\begin{itemize}
	\item $\pi_i=-c-\theta$ if  $i$ is vaccinated and is not infected;
	\item $\pi_i=-c-\theta-1$ if $i$ is vaccinated and is infected;
	\item $\pi_i=-1$ if $i$ is not vaccinated and is infected;
	\item $\pi_i=0$ if $i$ is not vaccinated and is not infected.
\end{itemize}
\end{center}
We investigate two scenarios for our extended model: one in which beliefs about vaccination, comprising vaccine-neutral and vaccine-averse attitudes, are fixed, and the other in which beliefs are contagious and concurrently spread in the same way as the imitation dynamics in vaccination decisions based on payoff differences. In this way, we explore systems in which both social contagion and epidemiological contagion are coupled, offering insight into the resulting disease-behavior system that exhibits dynamics not possible when the two subsystems are isolated from one another~\cite{fu2017dueling}.

\section*{Acknowledgements.}
Supported in part by a research grant from Investigator-Initiated Studies Program of Merck Sharp \& Dohme Corp. The opinions expressed in this paper are those of the authors and do not necessarily represent those of Merck Sharp \& Dohme Corp.

\end{document}